\documentclass[11pt]{article}
\usepackage{lscape}
\usepackage{graphicx}
\usepackage{amsmath}
\usepackage{amssymb}
\usepackage{hyperref}
\usepackage{graphicx,epic}
\usepackage{ulem}
\usepackage{multirow}
\usepackage{xypic}
\usepackage{wasysym}

\usepackage{array,multicol,color}

\definecolor{darkred}{rgb}{0.5,0,0.5}
\hypersetup{pdfborder={0 0 0},colorlinks=true,urlcolor=darkred,citecolor=blue,linkcolor=darkred}

\makeatletter

\@addtoreset{equation}{section}
\makeatother

\newcommand{\nn}{\nonumber}

\newcommand{\dlb}{[\![}
\newcommand{\drb}{]\!]}
\newcommand{\ad}{\textrm{ad\,}}

\newcommand{\h}{\mathfrak{h}}

\newcommand{\ints}{\mathbb{Z}}
\newcommand{\reals}{\mathbb{R}}

\newcommand{\bor}{\mathcal{B}}

\renewcommand{\em}{\it}

\begin{document}

\thispagestyle{empty}

\vskip-10pt
{\hfill{\tt AEI-2013-002}}\\
\vskip-10pt
{\hfill{\tt ULB-TH/12-22}}\\
\vskip-10pt
\hfill {\tt \today}\\
\vspace{20mm}

\begin{center} {\bf \LARGE Oxidizing Borcherds symmetries}

\vspace{15mm}

Axel Kleinschmidt${}^{1,\,2}$, Jakob Palmkvist${}^{3,\,4,\,2}$

\footnotesize
\vspace{9mm}

${}^1${\it Max Planck Institute for Gravitational Physics, Albert Einstein Institute\\
Am M\"uhlenberg 1, DE-14476 Potsdam, Germany\\[3mm]}
${}^2${\it  International Solvay Institutes\\ Campus Plaine C.P. 231, Boulevard du
Triomphe, BE-1050 Bruxelles, 
Belgium\\[3mm]}
${}^3${\it  Institut des Hautes Etudes Scientifiques\\ 35, Route de Chartres, FR-91440 Bures-sur-Yvette, 
France\\[3mm]}
${}^4${\it  Universit\'e Libre de Bruxelles\\ Campus Plaine C.P. 231, Boulevard du
Triomphe, BE-1050 Bruxelles, Belgium}

\vspace{24mm}

\hrule

\vspace{5mm}

\parbox{130mm}{

\noindent \footnotesize The tensor hierarchy of maximal supergravity in $D$ dimensions is known to be closely related to a Borcherds (super)algebra that is constructed from the global symmetry group $E_{11-D}$. We here explain how the Borcherds algebras in different dimensions are embedded into each other and can be constructed from a unifying Borcherds algebra. The construction also has a natural physical explanation in terms of oxidation. We then go on to show that the Hodge duality that is present in the tensor hierarchy has an algebraic counterpart. For $D> 8$ the Borcherds algebras we find differ from the ones existing in the literature although they generate the same tensor hierarchy.}

\vspace{5mm}

\hrule

\end{center}

\setcounter{equation}{0}
\setcounter{page}{1}

\newpage

\tableofcontents

\section{Introduction}

Maximal supergravity in $D\geq 2$ dimensions has a global hidden U-duality group with Lie algebra $E_{11-D}$~\cite{Cremmer:1978ds,Julia:1982gx,Nicolai:1987kz,Cremmer:1997ct}.
The theory admits propagating and non-propagating $p$-form potentials that transform in representations of this algebra in what is known as the tensor hierarchy~\cite{deWit:2005hv,Riccioni:2007au,Bergshoeff:2007qi,deWit:2008ta}.
Since all these algebras are embedded into each other, $E_{11-D} \subset E_{11-(D-1)}$, it is natural to consider them as subalgebras of the
infinite-dimensional Kac--Moody algebra $E_{11}$~\cite{West:2001as,Kleinschmidt:2003mf,Riccioni:2007au,Bergshoeff:2007qi} obtained when continuing to the extreme case $D=0$ (see also~\cite{Damour:2002cu,Bergshoeff:2008xv} for related work in the $E_{10}$ context).  
Based on the assumption that all U-dualities can be combined in $E_{11}$,
this gives a convenient unified description of all the tensor hierarchies in the various dimensions in that the hierarchies stem from various subalgebra decompositions of the adjoint of $E_{11}$ under its $\mathfrak{gl}(D)\oplus E_{11-D}$ subalgebra.

On the other hand, each of the finite-dimensional U-duality algebras $E_{11-D}$ for $D\geq 3$ can alternatively be extended to an infinite-dimensional so-called `V-duality' algebra~\cite{HenryLabordere:2002dk,Henneaux:2010ys}. This V-duality algebra admits a decomposition under the U-duality subalgebra and all the representations of the $p$-form potentials appear in this decomposition. The V-duality algebra is not a Kac--Moody algebra, but a generalization thereof known as a Borcherds (super)algebra~\cite{Borcherds,Ray}, which in turn is a special case of a contragredient Lie superalgebra \cite{Kac77A,Kac77B}.

The fact that the same representations appear in the level decompositions of both $E_{11}$ and the Borcherds algebra was explained in \cite{Henneaux:2010ys} (see also \cite{Palmkvist:2011vz,Palmkvist:2012nc}).
However, it should be stressed that $E_{11}$ contains representations that are not present (at the same level) in the Borcherds algebra, and the other way around. More precisely, the Kac--Moody algebra $E_{11}$ also has an infinity of tensor fields with mixed spacetime symmetry besides the antisymmetric fields~\cite{Kleinschmidt:2003mf,Riccioni:2006az,Riccioni:2007au,Bergshoeff:2007qi}, some of which can be turned into antisymmetric fields in lower dimensions by dimensional reduction. By contrast, the spectrum of the Borcherds V-duality algebra consists only of (antisymmetric) forms but of arbitrarily high rank; there is no upper limit from the spacetime dimension $D$. In fact, the space-time form-rank information is obtained by assigning an additional `V-degree' to the simple roots of the Borcherds algebra and this V-degree is then identified with the rank of the form in the tensor hierarchy. In the relation of~\cite{HenryLabordere:2002dk,Henneaux:2010ys} between the Borcherds algebra and $E_{11}$ the V-degree is associated with the tensor product of a parabolic subalgebra of $E_{11}$ and the outer form algebra in $D$ dimensions.

The interpretation of  the additional representations on each side is not clear so far and it remains to be seen whether one of the algebras is more likely than the other as a symmetry of M-theory. One advantage of the $E_{11}$ approach in this respect is its universality --- the same algebra can be used to derive the spectrum of $p$-form potentials for all $D$, whereas the Borcherds algebras are different for different $D$.

In this paper we will show that there is a similar universality also on the Borcherds side, and that
the V-duality algebras can be embedded into each other. This leads to the proposal of new V-duality algebras for $9 \leq D \leq 11$, different from the ones given in \cite{HenryLabordere:2002dk,Henneaux:2010ys}. We emphasize that these algebras lead to the same spectrum of $p$-form potentials, i.e., 
their `upper triangular' subalgebras are isomorphic, but there is no isomorphism 
when the whole algebras are considered. The embedding we study is physically motivated by the process of dimensional oxidation that allows us to identify which parts of an algebra have a higher-dimensional origin.

The paper is organized as follows. In section~\ref{sec:gkm}, we review Borcherds (super)algebras and which ones appear for maximal supergravity. We show that there is a natural embedding of the Borcherds algebras that arise in the various dimensions. The physical reason for this natural embedding is explained in more detail in section~\ref{sec:ox}. In section~\ref{sec:dual}, we discuss some more aspects of the symmetry algebras and their spectra, in particular in relation to Hodge duality.

\section{Chain of Borcherds algebras}
\label{sec:gkm}

After defining the concept of a Borcherds algebra in a way sufficient for our purposes, we will prove the main mathematical result of the paper: there exists a distinguished chain of Borcherds V-duality algebras that obey subalgebra relations. We give several different perspectives on this result from the mathematical side in this section. In section~\ref{sec:ox}, we reinterpret this result in physical terms.

\subsection{Borcherds preliminaries}

Like a Kac--Moody algebra~\cite{Kac}, a Borcherds algebra is uniquely defined by its Cartan matrix, which is a square matrix where each row and column corresponds to a simple root of the algebra~\cite{Borcherds,Ray}. However, the conditions that this matrix has to satisfy are less restrictive than in the Kac--Moody case so that Borcherds algebras constitute a true extension of the class of Kac--Moody algebras. In particular, they allow for the existence of imaginary simple roots.
The original Borcherds algebras defined in \cite{Borcherds} were further generalized to Borcherds superalgebras in \cite{Ray95}, allowing also for the existence of `fermionic' simple roots (the usual ones being `bosonic'). Borcherds superalgebras are in turn (in the case of finitely many simple roots) special cases of 
{\it contragredient} Lie superalgebras, defined already in \cite{Kac77A,Kac77B}.

Borcherds algebras are also called `generalized Kac--Moody' (GKM) algebras or `Borcherds--Kac--Moody' (BKM) algebras, but here we stick to the term `Borcherds algebras' for simplicity, and also use it for the Lie superalgebras generalizing the original Borcherds Lie algebras.

Given a Cartan matrix $A_{IJ}$ where $I$ and $J$ belong to a countable set of indices, one introduces so-called Chevalley generators $e_I$, $f_I$ and $h_I$ for each value of $I$.
Furthermore, one assigns a $\ints_2$-grading to the Chevalley generators $e_I$ and $f_I$, so that for each $I$, they are either both even (bosonic), or both odd (fermionic).

In order for the matrix $A_{IJ}$ to be a Cartan matrix of a Borcherds algebra it has to be real-valued and symmetric ($A_{IJ}=A_{JI}$), with non-positive off-diagonal entries ($A_{IJ} \leq 0$ if $I \neq J$) satisfying
\begin{align}
2\,\frac{A_{IJ}}{A_{II}}\in \mathbb{Z} \qquad  &\text{if}  \  A_{II} > 0, \nn\\  
\frac{A_{IJ}}{A_{II}} \in \mathbb{Z} \qquad  &\text{if}  \  A_{II} > 0  \  \text{and}  \   I \in S,
\end{align}
where $S$ is the set of indices $I$ such that $e_I$ and $f_I$ are odd. For simplicity we furthermore assume the Cartan matrix to be non-degenerate (otherwise one has to include additional semi-simple generators $h_a$). The restriction to symmetric matrices means that 
one does not necessarily have $A_{II}=2$ if $A_{II}>0$.

The Chevalley generators generate the Borcherds superalgebra $\bor$ subject to relations that we now specify, writing
the supercommutator of any two elements $x, y\in\bor$ as $\dlb x,y\drb$. One has $\dlb x,y\drb = \{ x,y\}=\{y,x\}$
if $x$ and $y$ are two odd elements
and $\dlb x,y\drb=[x,y]=-[y,x]$ if at least one of them is even.
The relations imposed on the generators of $\bor$ are
\begin{align} 
\dlb h_I,\,e_J \drb&=A_{IJ}e_J, & \dlb e_I,\,f_J \drb &=\delta_{IJ}h_J,\nn\\
\dlb h_I,\,f_J \drb&=-A_{IJ}f_J, & \dlb h_I,\,h_J \drb &=0, \label{chev-rel}
\end{align}
and the Serre relations
\begin{align} \label{serre-rel}
(\ad e_I)^{1-\frac{2A_{IJ}}{A_{II}}} (e_J) = (\ad f_I)^{1-\frac{2A_{IJ}}{A_{II}}} (f_J) &=0 \qquad \text{if}  \ \  A_{II}>0  \ \  \text{and} \ \   I \neq J,\nn\\
\dlb e_I,\,e_J \drb = \dlb f_I,\,f_J \drb &= 0 \qquad \text{if} \ \ A_{IJ}=0.
\end{align}

The $\ints_2$-grading on the Chevalley generators is extended to the whole of $\bor$ so that any supercommutator 
$\dlb x,y\drb$ is an even element if $x$ and $y$ have the same $\mathbb{Z}_2$-degree (odd/even), and an odd element if the
$\mathbb{Z}_2$-degrees are opposite.
As a consequence of this $\ints_2$-grading and the relation $\dlb e_I,f_J \drb =\delta_{IJ}h_J$, the generators $h_I$ are always even, and all the Lie supercommutators in (\ref{chev-rel}) involving $h_I$ can in fact be replaced by ordinary Lie commutators.

As for a Kac--Moody algebra 
the generators $h_I = \dlb e_I,\,f_I \drb $ span an abelian Cartan subalgebra $\h$ of $\bor$, and the dual space $\h^\ast$ 
is spanned by the
simple roots $\alpha_I$, defined by $\alpha_I (h_J) = A_{IJ}$.
An arbitrary element $\alpha$ in $\h^\ast$ is a root if there is an element
$e_\alpha$ in $\bor$ such that $[h_I,\,e_\alpha]=\alpha(h_I) e_\alpha$.
In particular
$e_{\alpha_I}=e_I$ and $e_{-\alpha_I}=f_I$ for the simple roots, which can consequently 
be divided into odd and even ones, with the
$\mathbb{Z}$-grading inherited from $\bor$.
The Cartan matrix defines a non-degenerate inner product on $\h^\ast$ by $(\alpha_I,\,\alpha_J)=A_{IJ}$, so that the diagonal value $A_{II}$ is the length (squared) of the simple root $\alpha_I$.
As for a Kac--Moody algebra we can also visualize the Cartan matrix with a Dynkin diagram, where $-A_{IJ}$ is the number of lines between two different nodes $I$ and $J$. But for a Borcherds algebra we also need to `paint' the nodes with different `colors', depending on the diagonal values $A_{II}$ and whether the corresponding simple roots are even or odd. Following \cite{HenryLabordere:2002dk,Henneaux:2010ys} we will here use white nodes for even simple roots of length (squared) 2, and black nodes for odd simple roots of zero length. (However, when we consider the more general contragredient Lie superalgebras in section \ref{subsec:non-dist}, we will switch to the convention of \cite{Kac77A,Kac77B} for the odd simple roots of zero length, representing them by `gray' nodes instead of black ones.) In cases where other types of simple roots appear, we will just write down the Cartan matrix instead of visualizing it with a Dynkin diagram (until we consider the contragredient Lie superalgebras in section \ref{subsec:non-dist}).

As mentioned in the introduction, one should also assign a V-degree to the simple roots of the Borcherds algebras that describe the
`V-duality' of maximal supergravity in $D$ dimensions. This assignment can then be extended to a linear map from the root space $\h^\ast$ to the set of non-negative integers. Defining it appropriately, all basis elements $e_\alpha$ of the Borcherds algebra,
where $\alpha$ is a root with V-degree $p$ ($1 \leq p \leq D$), transform in the same representation of the U-duality subalgebra as the $p$-form potentials of the theory~\cite{HenryLabordere:2002dk,Henneaux:2010ys}.

\subsection{Embedding of Borcherds algebras for $2 \leq D \leq 7$}

We now consider the specific Borcherds algebras that were introduced in~\cite{HenryLabordere:2002dk} (see also~\cite{Henneaux:2010ys}), starting with maximal supergravity in $D$ dimensions with $2 \leq D \leq 7$.
Here we call these algebras $\bor_n$, with $n=11-D$.
The U-duality algebra is $E_{n}$, and the 
Borcherds superalgebra $\bor_n$ that gives the tensor hierarchy is obtained by adding a single fermionic null root to the set of simple roots of 
$E_{n}$
with the resulting Dynkin diagram shown in Figure~\ref{Bn-dynkin-fig}, which also exhibits the labelling of nodes.
For this particular class of algebras $\bor_n$ the Serre relations (\ref{serre-rel}) simplify and reduce to
\begin{align}
\label{serre2}
(\ad e_I)^{1-A_{IJ}} (e_J) = (\ad f_I)^{1-A_{IJ}} (f_J) =0.
\end{align} 
for $A_{IJ}<0$.
The V-degree of the fermionic simple root $\alpha_0$ is equal to one, so that it generates a vector field ($1$-form). All other simple roots have V-degree zero~\cite{HenryLabordere:2002dk}.
\begin{center}
\begin{figure}[h]
\begin{picture}(450,75)(20,-10)
\put(115,-10){${\scriptstyle{0}}$}
\put(150,-10){${\scriptstyle{1}}$}
\put(205,-10){${\scriptstyle{n-4}}$}
\put(245,-10){${\scriptstyle{n-3}}$}
\put(285,-10){${\scriptstyle{n-2}}$}
\put(325,-10){${\scriptstyle{n-1}}$}
\put(260,45){${\scriptstyle{n}}$}
\thicklines
\multiput(210,10)(40,0){4}{\circle{10}}
\multiput(215,10)(40,0){3}{\line(1,0){30}}
\put(155,10){\circle{10}}
\put(115,10){\circle*{10}}
\put(120,10){\line(1,0){30}}
\multiput(160,10)(10,0){5}{\line(1,0){5}}
\put(250,50){\circle{10}} \put(250,15){\line(0,1){30}}
\end{picture} 
\caption{\it The Dynkin diagram of $\bor_n$.}
\label{Bn-dynkin-fig}
\end{figure}
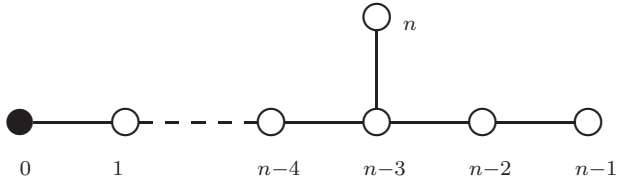
\end{center}

\subsection{The embedding $\bor_{n}\subset \bor_{n+1}$ for $3 \leq n \leq 8$} \label{subsec:emb} 

As we will have to refer to the Chevalley generators of both $\bor_n$ and $\bor_{n+1}$ in this subsection, we need to introduce different notations for them. The convention that we will use is to label the Chevalley generators of the larger $\bor_{n+1}$ with capital letters. The Chevalley generators of the embedded 
smaller $\bor_n$ will be denoted in turn by lowercase letters. 

Let $e_0$, $f_0$ and $h_0$  be given by
\begin{align}
e_0 &= \dlb E_0, E_1 \drb,& 
f_0 &= -\dlb F_0, F_1 \drb,& 
h_0 &= H_0 + H_1.
\end{align}
Define also for $i=1,\,\ldots,\,n$
\begin{align}
e_i &= E_{i+1},&
f_i &= F_{i+1},&
h_i &= H_{i+1},
\end{align}
so that the $E_{n}$ part of the diagram is inherited directly from $\bor_{n+1}$ to $\bor_n$. The generators $e_I$, $f_I$ and $h_I$ for $I=0,\,1,\,\ldots,\,n$ can be checked easily to be associated with the Cartan matrix of $\bor_n$. 
It is straightforward to show that the above generators $e_I$, $f_I$ and $h_I$ satisfy the defining relations (\ref{serre2}) of $\bor_n$. For instance, using the Jacobi superidentity one checks that
\begin{align}
\dlb e_0, e_0 \drb &= \dlb \dlb E_0, E_1 \drb, \dlb E_0, E_1 \drb\drb \nn\\
&= \dlb \dlb E_0, \dlb E_0, E_1\drb\drb,E_1\drb 
+\dlb E_0 \dlb E_1, \dlb E_0, E_1\drb\drb\drb 
=0,
\end{align} 
where the terms on the second line vanish by the Serre relations (\ref{serre2}) for $\bor_{n+1}$.
The other checks are similar.
This proves that $\bor_n$ is a subalgebra of $\bor_{n+1}$ for all $n\geq 3$. 
Another way to see this is to consider the root $\alpha_0+\alpha_1$ of $\bor_{n+1}$, with the labelling of Figure \ref{Bn-dynkin-fig}. It satisfies
\begin{align}
(\alpha_0+\alpha_1,\,\alpha_0+\alpha_1)&=0, & (\alpha_0+\alpha_1,\,\alpha_2)&=-1,
\end{align}
which shows that the root space of $\bor_n$ is a subspace of the root space of $\bor_{n+1}$. In the remaining cases we will only describe the embeddings in this way, and leave as an exercise for the reader to accordingly define the generators corresponding to the roots. This way of describing subalgebras is very similar to the one employed in~\cite{Feingold:2003es} for hyperbolic Kac--Moody algebras.

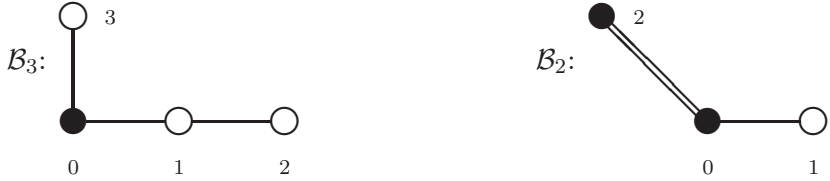
\begin{figure}[h]
\begin{center}
\begin{picture}(260,85)(65,-10)
\thicklines
\put(48,0){${\scriptstyle{0}}$}
\put(88,0){${\scriptstyle{1}}$}
\put(128,0){${\scriptstyle{2}}$}
\put(62,57){${\scriptstyle{3}}$}
\put(50,20){\circle*{10}}
\multiput(90,20)(40,0){2}{\circle{10}}
\multiput(55,20)(40,0){2}{\line(1,0){30}}
\put(50,60){\circle{10}} 
\put(50,25){\line(0,1){30}}
\put(25,40){$\bor_3$:}
\put(288,0){${\scriptstyle{0}}$}
\put(328,0){${\scriptstyle{1}}$}
\put(262,57){${\scriptstyle{2}}$}
\put(290,20){\circle*{10}}
\put(330,20){\circle{10}}
\put(295,20){\line(1,0){30}}
\put(250,60){\circle*{10}} 
\put(287,24){\line(-1,1){35}}
\put(286,22){\line(-1,1){35}}
\put(225,40){$\bor_2$:}
\end{picture} 
\caption{\label{B3-dynkin-fig}\it The Dynkin diagrams of $\bor_3$ (left) and $\bor_2$ (right).}
\end{center}
\end{figure}

The smallest algebra obtained by this construction is $\bor_3$, corresponding to $D=8$ maximal supergravity, and its Dynkin diagram is shown in 
Figure~\ref{B3-dynkin-fig} on the left.

\subsection{From $D=8$ to $D=9$: $\bor_2\subset \bor_3$} \label{subsec:8to9}

For $D=8$ the black node is attached to two white nodes in the Dynkin diagram (see Figure~\ref{B3-dynkin-fig}). From each of the two corresponding simple roots of length 2 we can obtain a null root by adding the simple null root corresponding to the black node. With the labelling of 
Figure~\ref{B3-dynkin-fig}
we thus have the two null roots $\alpha_0+\alpha_1$ and $\alpha_0+\alpha_3$. Together with $\alpha_2$, they can be considered as
simple roots of a subalgebra of $\bor_3$ which we call $\bor_2$. Its Cartan matrix is
\begin{align}
\label{b2cm}
\begin{pmatrix}
0 & -1 & -2\\
-1 & 2 & 0\\
-2 & 0 & 0 
\end{pmatrix}
\end{align}
and its Dynkin diagram is displayed in Figure~\ref{B3-dynkin-fig} on the right. The double line indicates that the entries in the off-diagonal corners of the Cartan matrix are equal to $-2$ instead of $-1$ as we would have for a single line, and thus that the scalar product of the corresponding simple roots is equal to $-2$. Indeed, one checks easily using the embedding in $\bor_3$ that
\begin{align}
(\alpha_0+\alpha_1,\,\alpha_0+\alpha_3)= (\alpha_0,\,\alpha_1)+(\alpha_0,\,\alpha_3) =-2.
\end{align}
The V-degree of both fermionic simple roots is equal to one. 

The Cartan matrix (\ref{b2cm}) is different from the one proposed in~\cite{HenryLabordere:2002dk,Henneaux:2010ys} which has inner product $-1$ between the two fermionic simple roots. Due to the Chevalley--Serre relations (\ref{chev-rel})--(\ref{serre-rel}) this does not lead to any difference in the structure of the algebra in the positive triangular part generated by the $e_I$. Therefore the spectrum of $p$-forms is identical in both cases. However, the algebra $\bor_2$ of (\ref{b2cm}) and the one of~\cite{HenryLabordere:2002dk} are not isomorphic when the remaining generators (Cartan and negative triangular) are taken into account.

\subsection{From $D=9$ to $D=10$, type IIA and type IIB}

When going from $D=9$ to $D=10$ there are two choices associated with type IIA and type IIB supergravity. We start with type IIB.

One combines the two simple fermionic null roots of $\bor_2$ into a bosonic root of length $-4$. Using the labelling of Figure~\ref{B3-dynkin-fig} for $\bor_2$ we have:
\begin{align}
(\alpha_0+\alpha_2,\,\alpha_0+\alpha_2)=2(\alpha_2,\,\alpha_0)=-4.
\end{align}
Taking this as a simple root of a subalgebra $\bor_{1{\rm B}}$, together with $\alpha_1$,
the resulting Cartan matrix is
\begin{align}
\begin{pmatrix}
-4 & -1\\
-1 & 2
\end{pmatrix}
\end{align}
and this is now a Borcherds Lie algebra, not a proper Borcherds Lie {superalgebra}. The V-degree of 
the simple root of negative length, which is the root $\alpha_0+\alpha_2$ of $\bor_2$, is equal to 2, the sum of the V-degrees of
$\alpha_0$ and $\alpha_2$ in $\bor_2$.

For type IIA we again combine the two simple fermionic null roots into a bosonic root of length $-4$, 
but we also combine $\alpha_0$ and $\alpha_1$ into
a simple fermionic null root of a subalgebra $\bor_{1{\rm A}}$. The reason for this will be clear in the next section when we discuss the relation to oxidation. Using the labelling of Figure~\ref{B3-dynkin-fig} for $\bor_2$ we end up with the scalar product
\begin{align}
(\alpha_0+\alpha_1,\,\alpha_0+\alpha_2)=-3
\end{align}
and the Cartan matrix
\begin{align}
\begin{pmatrix}
-4 & -3\\
-3 & 0
\end{pmatrix}.
\end{align}
These algebras are again not isomorphic to the ones in~\cite{HenryLabordere:2002dk} but agree on the positive triangular part. The V-degrees of the two simple generators are $2$ and $1$, corresponding to the two-form and vector field of type IIA.

\subsection{From $D=10$ to $D=11$}
\label{subsec:10to11}

Here we combine the two simple roots of the type IIA algebra into a simple fermionic root of length $-10$. The Borcherds algebra is finite-dimensional in this case, and isomorphic to the one given in \cite{HenryLabordere:2002dk}, where the simple fermionic root has length $-1$. Since the Cartan matrix has only one single entry (and this entry is non-zero), this is just a matter of normalisation, and both Borcherds algebras coincide with the 5-dimensional Lie superalgebra $\mathfrak{osp}(1|2)$.
The V-degree of the only simple root is equal to $3$ (the sum of the V-degrees of the simple roots of $\bor_{1{\rm A}}$), corresponding to the three-form of eleven-dimensional supergravity.

\subsection{Decomposition of the representations}
\label{sec:repdec}

The construction above can be understood also in the following way. The (adjoint of the) superalgebra $\bor_{n+1}$ has a level decomposition with respect to node $0$ from the following subalgebra
\begin{align}
E_{n+1}\oplus\reals\subset\bor_{n+1}
\end{align}
which arranges the $\bor_{n+1}$ generators in $E_{n+1}$ representations as
\begin{align}
\label{pdec}
\bor_{n+1} = \bigoplus_{p\in\ints} s_p.
\end{align}
Here $s_p$ is related to the space of $p$-forms in $10-n$ dimensions (for $p\leq 10-n$). The space $s_p$ is odd/even when $p$ is odd/even and is a (finite-dimensional) representation of $E_{n+1}$. 

We can also perform a $\ints^2$-graded decomposition of $\bor_{n+1}$ corresponding to 
\begin{align}
E_n\oplus \reals \oplus\reals\subset \bor_{n+1},
\end{align}
that is, a double level decomposition with respect to nodes $0$ and $1$. (For $n<2$, the grading has to be adapted but the results below still hold.)
Associated with this double grading one obtains a graded decomposition of $s_p$:
\begin{align}
s_p = \bigoplus_{q\in\ints} s_{p,q},
\end{align}
where we have normalised the grading generator such that the charges are integral. (The $q$-sum is only finite, but we write it in this more general form for simplicity.)
The double grading of the superalgebra $\bor_{n+1}$ can be written as:
\begin{align}
\bor_{n+1} =\bigoplus_{p,q} s_{p,q},\quad\quad \dlb s_{p,q}, s_{p',q'} \drb = s_{p+p',q+q'}.
\end{align}
All $s_{p,q}$ are representations of $E_n$. If we now restrict to the `diagonal' spaces $s_{p,p}$, i.e. $p=q$, we can study the superalgebra
\begin{align}
\label{diagsub}
\bigoplus_{p\in\ints} s_{p,p} \subset \bor_{n+1}.
\end{align}
This is a subalgebra of $\bor_{n+1}$ and is exactly the algebra that is generated by the simple generators that were defined above. We conclude that
\begin{align}
\bor_n=\bigoplus_{p\in\ints} s_{p,p} \subset \bor_{n+1}.
\end{align}

As all the Borcherds superalgebras can be embedded into one another, one might work simply with a large one, say $\bor_{11}$, with Dynkin diagram as in Figure \ref{Bn-dynkin-fig} with $n=11$, that generates all the $p$-form hierarchies in all dimensions. With this we mean that $\bor_{11}$ contains the subalgebra $\bor_n$ that generates the $p$-form hierarchy in $D=11-n$ dimensions. The $p$-form hierarchy is obtained from the space $s_p$ in the standard way by attaching to each $s_p$ the V-degree as form rank~\cite{Henneaux:2010ys}.

\subsection{Non-distinguished Dynkin diagrams}
\label{subsec:non-dist}

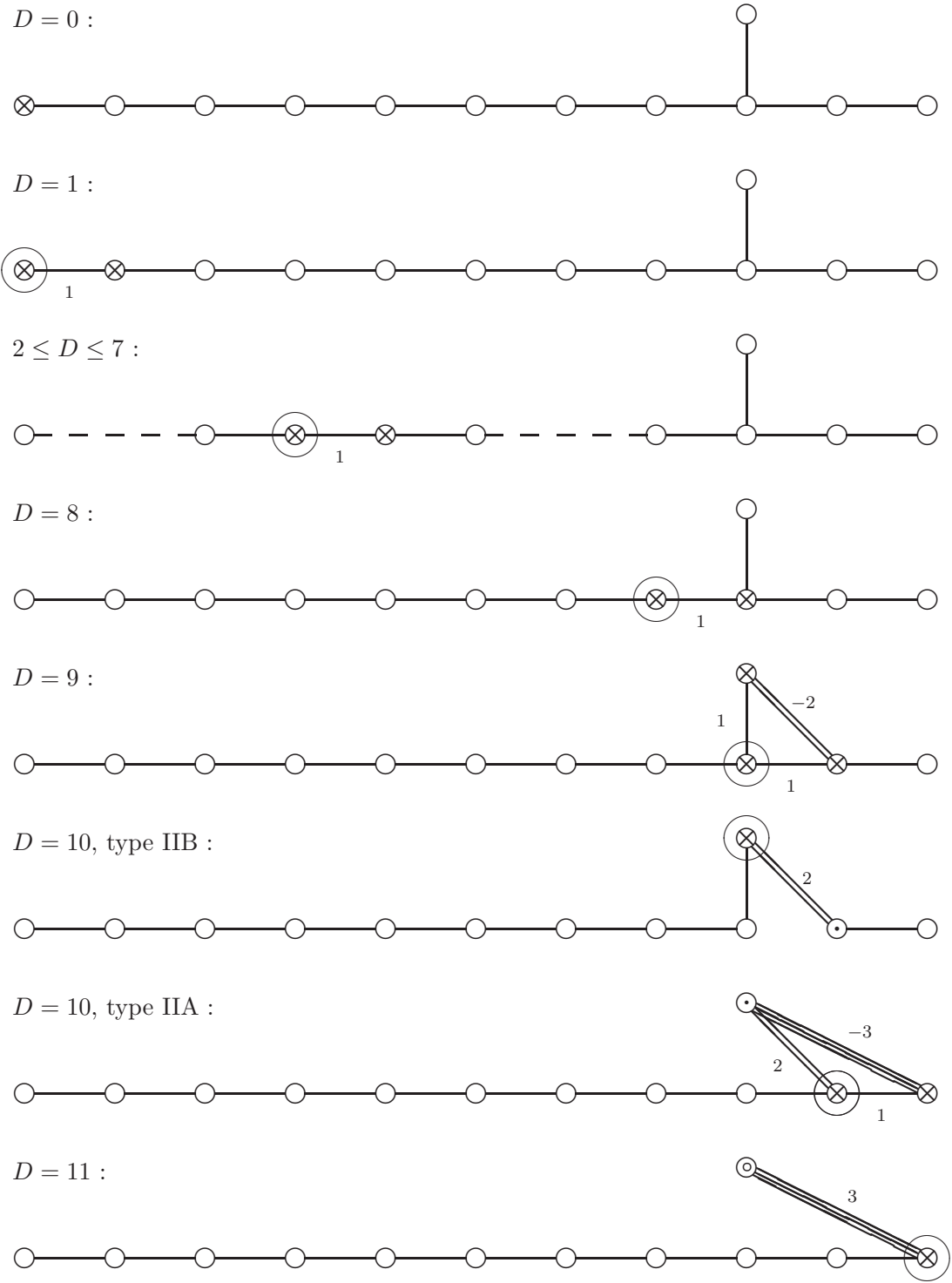
\begin{figure} 
\begin{center}
\scalebox{0.98}{
\begin{picture}(450,72)(-67,-10)
\thicklines
\put(-70,45){$D=0$ :}
\multiput(-70,7.165)(40,0){11}{$\mbox{\boldmath {\large $\ocircle$ }}$}
\put(-70,7.165){$\mbox{\boldmath {\large $\times$ }}$}
\multiput(-60.5,10)(40,0){10}{\line(1,0){31.5}}
\put(250,47.5){$\mbox{\boldmath {\large $\ocircle$ }}$}
\put(255.3,14.5){\line(0,1){31.5}}
\end{picture}
}
\scalebox{0.98}{
\begin{picture}(450,72)(-67,-10)
\thicklines
\put(-70,45){$D=1$ :}
\thinlines
\put(-64.7,10.2){\circle{20}}
\thicklines
\multiput(-70,7.165)(40,0){11}{$\mbox{\boldmath {\large $\ocircle$ }}$}
\put(-70,7.165){$\mbox{\boldmath {\large $\times$ }}$}
\put(-30,7.165){$\mbox{\boldmath {\large $\times$ }}$}
\multiput(-60.5,10)(40,0){10}{\line(1,0){31.5}}
\put(250,47.5){$\mbox{\boldmath {\large $\ocircle$ }}$}
\put(255.3,14.5){\line(0,1){31.5}}
\put(-47,-2){\scriptsize $1$}
\end{picture} 
}
\scalebox{0.98}{
\begin{picture}(450,72)(-67,-10)
\thicklines
\put(-70,45){$2 \leq D \leq 7$ :}
\thinlines
\put(55.3,10.2){\circle{20}}
\thicklines
\multiput(10,7.165)(40,0){4}{$\mbox{\boldmath {\large $\ocircle$ }}$}
\multiput(210,7.165)(40,0){4}{$\mbox{\boldmath {\large $\ocircle$ }}$}
\put(-70,7.165){$\mbox{\boldmath {\large $\ocircle$ }}$}
\put(50,7.165){$\mbox{\boldmath {\large $\times$ }}$}
\put(90,7.165){$\mbox{\boldmath {\large $\times$ }}$}
\multiput(19.5,10)(40,0){3}{\line(1,0){31.5}}
\put(-60.7,10){\line(1,0){8}}
\put(-44.7,10){\line(1,0){8}}
\put(-28.7,10){\line(1,0){8}}
\put(-12.7,10){\line(1,0){8}}
\put(3.3,10){\line(1,0){8}}
\put(139.4,10){\line(1,0){7.9}}
\put(155.3,10){\line(1,0){8}}
\put(171.3,10){\line(1,0){8}}
\put(187.3,10){\line(1,0){8}}
\put(203.3,10){\line(1,0){7.9}}
\multiput(219.5,10)(40,0){3}{\line(1,0){31.5}}
\put(250,47.5){$\mbox{\boldmath {\large $\ocircle$ }}$}
\put(255.3,14.5){\line(0,1){31.5}}
\put(73,-2){\scriptsize $1$}
\end{picture}
}
\scalebox{0.98}{
\begin{picture}(450,72)(-67,-10)
\thicklines
\put(-70,45){$D=8$ :}
\thinlines
\put(215.3,10.2){\circle{20}}
\thicklines
\multiput(-70,7.165)(40,0){11}{$\mbox{\boldmath {\large $\ocircle$ }}$}
\put(210,7.165){$\mbox{\boldmath {\large $\times$ }}$}
\put(250,7.165){$\mbox{\boldmath {\large $\times$ }}$}
\multiput(-60.5,10)(40,0){10}{\line(1,0){31.5}}
\put(250,47.5){$\mbox{\boldmath {\large $\ocircle$ }}$}
\put(255.3,14.5){\line(0,1){31.5}}
\put(233,-2){\scriptsize $1$}
\end{picture}
}
\scalebox{0.98}{
\begin{picture}(450,72)(-67,-10)
\thicklines
\put(-70,45){$D=9$ :}
\thinlines
\put(255.3,10.2){\circle{20}}
\thicklines
\multiput(-70,7.165)(40,0){11}{$\mbox{\boldmath {\large $\ocircle$ }}$}
\put(250,7.165){$\mbox{\boldmath {\large $\times$ }}$}
\put(250,47.5){$\mbox{\boldmath {\large $\times$ }}$}
\put(290,7.165){$\mbox{\boldmath {\large $\times$ }}$}
\multiput(-60.5,10)(40,0){10}{\line(1,0){31.5}}
\put(250,47.5){$\mbox{\boldmath {\large $\ocircle$ }}$}
\put(255.3,14.5){\line(0,1){31.5}}
\put(291.4,12.7){\line(-1,1){33.8}}
\put(275,35){\scriptsize $-2$}
\put(273,-2){\scriptsize $1$}
\put(242,27){\scriptsize $1$}
\put(293.2,14.2){\line(-1,1){33.8}}
\end{picture} 
}
\scalebox{0.98}{
\begin{picture}(450,72)(-67,-10)
\thicklines
\put(-70,45){$D=10$, type IIB :}
\thinlines
\put(255.3,50.5){\circle{20}}
\thicklines
\multiput(-70,7.165)(40,0){11}{$\mbox{\boldmath {\large $\ocircle$ }}$}
\put(250,47.5){$\mbox{\boldmath {\large $\times$ }}$}
\put(293.5,7.165){$\mbox{\boldmath {\large $\cdot$ }}$}
\multiput(-60.5,10)(40,0){8}{\line(1,0){31.5}}
\put(299.5,10){\line(1,0){31.5}}
\put(280,30){\scriptsize $2$}
\put(250,47.5){$\mbox{\boldmath {\large $\ocircle$ }}$}
\put(255.3,14.5){\line(0,1){31.5}}
\put(291.4,12.7){\line(-1,1){33.8}}
\put(293.2,14.2){\line(-1,1){33.8}}
\end{picture}
}
\scalebox{0.98}{
\begin{picture}(450,72)(-67,-10)
\thicklines
\put(-70,45){$D=10$, type IIA :}
\thinlines
\put(295.3,10.2){\circle{20}}
\thicklines
\put(313,-2){\scriptsize $1$}
\put(267,20){\scriptsize $2$}
\multiput(-70,7.165)(40,0){11}{$\mbox{\boldmath {\large $\ocircle$ }}$}
\put(290,7.165){$\mbox{\boldmath {\large $\times$ }}$}
\put(330,7.165){$\mbox{\boldmath {\large $\times$ }}$}
\multiput(-60.5,10)(40,0){10}{\line(1,0){31.5}}
\put(250,47.5){$\mbox{\boldmath {\large $\ocircle$ }}$}
\put(250,47.6){$\mbox{\boldmath { $\cdot$ }}$}
\put(291.4,12.7){\line(-1,1){33.8}}
\put(293.2,14.2){\line(-1,1){33.8}}
\put(330.7,10.5){\line(-2,1){72.5}}
\put(330.7,12.5){\line(-2,1){71.5}}
\put(330.7,14.5){\line(-2,1){70.8}}
\put(269.3,41.2){\line(2,-1){30.5}}
\put(269.3,43.2){\line(2,-1){62}}
\put(269.3,45.2){\line(2,-1){63.2}}
\put(300,35){\scriptsize $-3$}
\thinlines
\put(295.3,10.2){\circle{20}}
\end{picture} 
}
\scalebox{0.98}{
\begin{picture}(450,52)(-67,10)
\put(-70,45){$D=11$ :}
\thinlines
\put(335.3,10.2){\circle{20}}
\thicklines
\put(300,35){\scriptsize $3$}
\multiput(-70,7.165)(40,0){11}{$\mbox{\boldmath {\large $\ocircle$ }}$}
\put(249.4,48.5){$\mbox{\boldmath { \scriptsize $\circ$ }}$}
\put(330,7.165){$\mbox{\boldmath {\large $\times$ }}$}
\multiput(-60.5,10)(40,0){10}{\line(1,0){31.5}}
\put(250,47.5){$\mbox{\boldmath {\large $\ocircle$ }}$}
\put(330.7,10.5){\line(-2,1){72.5}}
\put(330.7,12.5){\line(-2,1){71.5}}
\put(330.7,14.5){\line(-2,1){70.8}}
\put(269.3,41.2){\line(2,-1){30.5}}
\put(269.3,43.2){\line(2,-1){62}}
\put(269.3,45.2){\line(2,-1){63.2}}
\end{picture}
}
\caption{\it Different Dynkin diagrams of $\bor_{11}$. See the main text for explanations. \label{non-dist-fig}}
\end{center}
\end{figure}

The embeddings that we have described are not obvious from the Dynkin diagrams, in contrast to the embeddings $E_n \subset E_{n+1}$, where one just has to remove a node from the Dynkin diagram of $E_{n+1}$ to obtain the one of $E_n$. However, 
considered as special cases of contragredient Lie superalgebras,
the Dynkin diagrams of the Borcherds algebras are not unique. In this subsection\footnote{For these results, we benefitted greatly from discussions with Bernard Julia and Victor Kac.} we will show that one can in fact choose the Dynkin diagrams of $\bor_n$ and $\bor_{n+1}$ such that the embedding $\bor_n \subset \bor_{n+1}$ becomes manifest, just as $E_n \subset E_{n+1}$. 
These new Dynkin diagrams can be obtained from those in Figure \ref{Bn-dynkin-fig} and \ref{B3-dynkin-fig} by applying so-called generalized Weyl reflections that 
transform the set of simple roots into a new one. 
A generalized Weyl reflection is associated to an odd simple null root $\alpha_I$ ($I \in S,\,A_{II}=0$),
and acts on the simple roots by ($J\neq I$)
\begin{subequations}
\begin{align}
\alpha_{I}& \mapsto -\alpha_{I},&&\\
\alpha_J &\mapsto \left\{\begin{array}{rl}\alpha_J + \alpha_I,&\textrm{if $A_{IJ}\neq 0$},\\
\alpha_J, &\textrm{if $A_{IJ}= 0$}.\end{array}\right.
\end{align}
\end{subequations}
Applying this repeatedly to $\bor_{11}$, with a set of simple roots corresponding to the Dynkin diagram in Figure~\ref{Bn-dynkin-fig} (with $n=11$), one obtains different sets of simple roots
corresponding to
different Dynkin diagrams, as illustrated in Figure \ref{non-dist-fig}. The first Dynkin diagram is the distinguished one in the sense that there
is only one odd simple null root, and the others are non-distinguished Dynkin diagrams of $\bor_{11}$.

Generalized Weyl reflections do not preserve 
the inner product on the root space $\h^\ast$ (unlike the standard Weyl reflections)
and can thus transform a standard Cartan matrix into one with positive off-diagonal entries. Such a matrix does not satisfy the requirements for a Cartan matrix of a Borcherds algebra, but it still defines a contragredient Lie superalgebra \cite{Kac77A,Kac77B}.
Although we do not change the Borcherds algebra, it is thus convenient to go to the more general class of contragredient Lie superalgebras. In the figure we therefore switch to the convention of \cite{Kac77A,Kac77B} for the coloring of the nodes associated to odd simple roots of zero length.
Instead of black nodes, we use `gray' ones ($\otimes$). This is summarized below, where we also introduce two more types of nodes for the simple roots that we encounter:
\setlength{\arraycolsep}{5pt}
{\renewcommand{\arraystretch}{1.5}
\begin{align*}
\begin{array}{cl}
\ocircle & \text{even root of length 2 ($I \notin S,\ A_{II}=2$)},\nn\\
\otimes  & \text{odd root of length 0 ($I \in S,\ A_{II}=0$)}\nn\\
& \text{(previously represented 
by a black node)},\nn\\
\odot & \text{even root of length $-4$ ($I \notin S,\ A_{II}=-4$)},\nn\\
{ \circledcirc} & \text{odd root of length $-10$ ($I \in S,\ A_{II}=-10$)}.
\end{array}
\end{align*}}

\noindent

The number of lines between two nodes $I$ and $J$ is still related to off-diagonal entry $A_{IJ}$ in the Cartan matrix, but it is not always equal to $-A_{IJ}$ as for a Borcherds algebra. Rather, it is equal to $|A_{IJ}|$ --- 
since the off-diagonal entries can be positive, the number of lines between the nodes only determine them up to a sign. However, as long as
one of the nodes $I$ and $J$ is white, $A_{IJ}$ is still negative (in the diagrams that we consider), and in the remaining cases we have written out $A_{IJ}$ explicitly next to the lines to fix the ambuiguity.

We stress that all these diagrams describe the same algebra, namely $\bor_{11}$. But the same procedure can be applied also to the Borcherds algebras $\bor_{n}$ with $n < 11$, and we will end up with a diagram without any branching, like the ones in Figure \ref{non-dist-fig} for $\bor_{11}$ corresponding to $D=11$ or $D=10$, type IIB, but with fewer white nodes. Then it becomes obvious that $\bor_n \subset \bor_{n+1}$: we can obtain the Dynkin diagram of $\bor_n$ from that of $\bor_{n+1}$ by just removing the leftmost white node.

This way of analysing the algebra $\bor_{11}$ also makes an easier contact to the space-time structure of the generators: As indicated in the figure, the diagrams can be associated to maximal supergravity in $D$ dimensions, $0 \leq D \leq 11$. In each diagram there is an odd simple null root such that its removal leads to two diagrams corresponding to $\mathfrak{sl}(D)$ on the left hand side, and $\bor_{11-D}$ on the right hand side (for $D\geq 2$). We have put a circle around this node, which is also associated to the generalized Weyl reflection that `reduces' the diagram from $D$ to $D-1$ dimensions. Thus one can see that there is an algebra $\mathfrak{sl}(D)$ commuting with $\bor_{11-D}$ within $\bor_{11}$. This is very similar to the way one obtains the forms in $D$ dimensions from $E_{11}$ by decomposing the adjoint representation of $E_{11}$ into representations of $\mathfrak{sl}(D)\oplus E_{11-D}$~\cite{Kleinschmidt:2003mf,Riccioni:2007au,Bergshoeff:2007qi}.

\section{Relation to decompactification/oxidation}
\label{sec:ox}

The algebraic construction of the embedding of the various Borcherds algebras has a simple physical counterpart in terms of Kaluza--Klein reduction and oxidation of maximal ungauged supergravity. Let the $D$-dimensional metric come from a circle reduction of a $(D+1)$-dimensional theory as in
\begin{align}
ds_{D+1}^2 = e^{2\alpha \phi} ds_D^2 + e^{2\beta\phi} \left( dz + A_\mu dx^\mu \right)^2,
\end{align}
where $z$ is the circle direction, $\phi$ the dilaton and $A_\mu$ the Kaluza--Klein vector. The exponents
\begin{align}
\alpha = -\frac{1}{\sqrt{2(D-1)(D-2)}},\quad \beta = \sqrt{\frac{D-2}{2(D-1)}}
\end{align}
are chosen such that one reduces from Einstein frame to Einstein frame with a canonically normalised scalar field $\phi$. The (expectation value of the) dilaton is related to the radius $R_{D+1}$ of the circle in the direction $D+1$ via
\begin{align}
e^{\langle \beta\phi\rangle} = \frac{R_{D+1}}{\ell_{D+1}},
\end{align}
where $\ell_{D+1}$ is the $(D+1)$-dimensional Planck length.
The scalar field $\phi$ parametrises a particular direction in the Cartan subalgebra of $E_{11-D}$ that is the symmetry of the reduced theory. The $p$-forms in $D$ dimensions transform in a representation of $E_{11-D}$ and they carry a particular weight under this direction in the Cartan subalgebra. In fact, the direction corresponds to the decomposition
\begin{align}
E_{10-D} \oplus \reals \subset E_{11-D}.
\end{align}
It is now important that the power of the dilaton also depends on the form degree and so it is a combination of the eigenvalue of the fundamental weight associated with the $\reals$ summand in the decomposition above and the form degree. This exactly parallels the discussion of the double gradation in section (\ref{sec:repdec}). More precisely, a $p$-form potential reduces from $D+1$ to $D$ dimensions via
\begin{align}
|F_{p+1}|^2 \to e^{-2\alpha(p+1)\phi} |F_{p+1}|^2 +(p+1) e^{-2(\alpha p +\beta)\phi} |F_{p}|^2,
\end{align}
where we have written everything in terms of the field strengths and the right hand side contains the fields in $D$ space-time dimensions.

The key point is now that in oxidation from $D$ to $D+1$ dimensions, the terms with the largest powers of $(R_{D+1}/\ell_{D+1})$ are dominant and are the only ones that survive. These are the ones to be kept in the decompactification process and is a condition on the dilaton power. When normalising the $\reals$ factor conveniently, the powers become integral. We illustrate this by some examples.

In $D=3$ one obtains the decomposition of the various fields in the hierarchy as displayed in Table~\ref{3dTH}.
We have underlined in all cases the pieces where the $\reals$-charge $q$ is equal to the $p$-form rank, according to condition (\ref{diagsub}). In all cases except for $p=1$, this piece also corresponds to the lowest charge that is available for a given value of $p$. Therefore these terms are the ones that decompactify to $D+1$ dimensions to give $p$-forms there. The reason that there is one additional more dominant singlet vector ($p=1$) is that it comes from the Kaluza--Klein vector of the metric reduction and also oxidises to the higher-dimensional metric. With this reasoning one obtains exactly the right fields from oxidation and one has the same condition that the $\reals$-charge $q$ equals the form rank as in the algebraic construction~(\ref{diagsub}). In principle, the Borcherds algebra $\bor_8$ also predicts representations with $p=4$ that would be interpreted as four-forms in three dimensions. We will come back to these representations in section~\ref{sec:BY}.

As another example we consider the oxidation of the tensor hierarchy from eight to nine dimensions. For $D=8$ one has the U-duality symmetry algebra $E_3=\mathfrak{sl}(3)\oplus\mathfrak{sl}(2)$ whereas the $D=9$ algebra is $E_2=\mathfrak{gl}(2)=\mathfrak{sl}(2)\oplus \reals$. We label the representations of $E_2$ as ${\mathbf n}_h$ where $n$ is the dimension of an irreducible $\mathfrak{sl}(2)$ representation and $h$ is the eigenvalue under the direct $\reals$ summand that appears in $E_2$ for which we choose the normalisation that the vector doublet is a genuine doublet. (In most literature on the subject this charge is not given explicitly.) The $q$-grade is defined by the sum of the fundamental weights of nodes $1$ and $3$ in Figure~\ref{B3-dynkin-fig}. With this we obtain Table~\ref{8dTH}.

\begin{landscape}
{\renewcommand{\arraystretch}{1.5}
\begin{table}
\begin{tabular}
{|c||c||c|c|c|c|c|c|c|c|c|c|}
\hline
$p$ & $s_p$ & $q=0$ & $q=1$ & $q=2$ & $q=3$ & $q=4$ & $q=5$ & $q=6$ & $q=7$ & $q=8$ & $q=9$\\
\hline\hline
$1$ & ${\bf 248}$ & ${\bf 1}$ & \underline{${\bf 56}$} & ${\bf 133}\,\oplus\, {\bf 1}$ & ${\bf 56}$ & ${\bf 1}$&&&&&\\
\hline
\multirow{2}{*}{$2$} & ${\bf 3875}$ &&&  \underline{${\bf 133}$} & ${\bf 912}\,\oplus\,{\bf 56}$ & 
$\begin{matrix}{\bf 1539}\,{\oplus}\\\,{\bf 133}
\,{\oplus\,{\bf 1}}\end{matrix}$ & ${\bf 912}\,\oplus\,{\bf 56}$ & ${\bf 56}$ &&&\\\cline{2-12}
& ${\bf 1}$ &&&&& ${\bf 1}$ &&&&&\\\hline
\multirow{3}{*}{$3$} & ${\bf 147\,250}$ &&&& \underline{${\bf 912}$} & $\begin{matrix}{\bf 8645}\\\oplus\,{\bf 1539}\\\oplus\,{\bf 133}
\end{matrix}$ & $\begin{matrix}{\bf27\,664}\\\oplus\,{\bf 6480}\\\oplus\, {2\times{\bf 912}}\\\oplus\, {\bf 56}\end{matrix}$ & 
$\begin{matrix}{\bf 40\,755}\\\oplus\,{\bf 8645}\\{\oplus\,2\times{\bf 1539}}\\\oplus\,2\times{\bf 133}\end{matrix}$ &$\begin{matrix}{\bf27\,664}\\\oplus\,{\bf 6480}\\\oplus\, {2\times{\bf 912}}\\\oplus\, {\bf 56}\end{matrix}$ & $\begin{matrix}{\bf 8645}\\\oplus\,{\bf 1539}\\\oplus\,{\bf 133}
\end{matrix}$ & ${\bf 912}$\\\cline{2-12}
& ${\bf 3875}$ &&&&&  ${\bf 133}$ & ${\bf 912}\,\oplus\,{\bf 56}$ & $\begin{matrix}{\bf 1539}\,{\oplus}\\\,{\bf 133}
\,{\oplus\,{\bf 1}}\end{matrix}$ & ${\bf 912}\,\oplus\,{\bf 56}$ & ${\bf 133}$ &\\\cline{2-12}
& ${\bf 248}$ &&&&& ${\bf 1}$ & ${\bf 56}$ & ${\bf 133}\oplus {\bf 1}$ & ${\bf 56}$ & ${\bf 1}$&\\\hline
\end{tabular}
\caption{\label{3dTH}\it Hierarchy of $p$-form fields as predicted by the Borcherds algebra $\bor_8$. The column $s_p$ lists the $E_8$ representations of the $p$-forms (to be precise, $s_p$ is the direct sum of the irreduciple representations given for each $p$). The following columns contain the decomposition under $E_7\oplus \reals\subset E_8$. The label $q$ is related to the charge under the summand $\reals$ as explained in section~\ref{sec:repdec}. The hierarchy predicted by $\bor_8$ for $p>3$ is also non-empty but not displayed here. We discuss some aspects of it in section~\ref{sec:BY}.}
\end{table}
}
\end{landscape}

\setlength{\arraycolsep}{-100pt}
{\renewcommand{\arraystretch}{1.5}
\begin{table}[h!]
\begin{tabular}{|c||c||c|c|c|c|c|c|c|c|}
\hline
$p$ & $s_p$ & $q=0$ & $q=1$ & $q=2$ & $q=3$ & $q=4$ & $q=5$ & $q=6$ & $q=7$ \\
\hline\hline
$1$ & $({\bf 3},{\bf 2})$ &  ${\bf 1}_0$ & $\!$\underline{${\bf 2}_1\oplus {\bf 1}_{-4/3}$}$\!$ & ${\bf 2}_{-1/3}$ &&&&&\\\hline
$2$ & $({\bf \overline{3}}, {\bf 1})$ &&&\underline{${\bf 2}_{-1/3}$} & ${\bf 1}_{2/3}$ &&&&\\\hline
$3$ & $({\bf 1},{\bf 2})$ &&&& \underline{${\bf 1}_{2/3}$} & ${\bf 1}_{-2/3}$ &&&\\\hline
$4$ & $({\bf 3},{\bf 1})$ &&&&&  \underline{${\bf 1}_{-2/3}$} & ${\bf 2}_{1/3}$ &&\\\hline
$5$ & $(\overline{\bf 3},{\bf 2})$ &&&&&&  \underline{${\bf 2}_{1/3}$} & $\!{\bf 2}_{-1}\oplus {\bf 1}_{4/3}\!$ & ${\bf 1}_0$\\\hline
\end{tabular}
\caption{\label{8dTH}\it Hierarchy of $p$-form fields in $D=8$ as predicted by the Borcherds algebra $\bor_3$. The column $s_p$ lists the $E_3=\mathfrak{sl}(3)\oplus\mathfrak{sl}(2)$ representation and the columns with different $q$ represent the decomposition under $E_2=\mathfrak{sl}(2)\oplus\reals$. The notation ${\bf n}_h$ denotes the $n$-dimensional representation of $\mathfrak{sl}(2)$ with eigenvalue $h$ under the $\reals$ summand. The underlined representations are the ones that survive the oxidation process (besides the Kaluza--Klein vector). We have truncated the table at $p=5$.}
\end{table}
}

\setlength{\tabcolsep}{6.3pt}
{\renewcommand{\arraystretch}{1.5}
\begin{table}[p]
\begin{tabular}{|c||c||c|c|c|c|c|c|c|c|c|}
\hline
$p$ & $s_p$ & $q=0$ & $q=1$ & $q=2$ & $q=3$ & $q=4$ & $q=5$ & $q=6$ & $q=7$ & $q=8$\\
\hline\hline
\multirow{2}{*}{$1$} & ${\bf 2}_1$ 
&  ${1}$ & 
$\underline{1}$ &&&&&&&\\\cline{2-11}
&${\bf 1}_{-4/3}$ &  & & $1$ &&&&&&\\\hline
$2$ & ${\bf 2}_{-1/3}$ &&& \underline{$1$} & $1$ &&&&&\\\hline
$3$ & ${\bf 1}_{2/3}$ &&&& \underline{$1$} & &&&&\\\hline
$4$ & ${\bf 1}_{-2/3}$ &&&&&& $1$ &&&\\\hline
$5$ & ${\bf 2}_{1/3}$ &&&&&&  \underline{$1$} & $1$ &&\\\hline
\multirow{2}{*}{$6$} & ${\bf 1}_{2/3}$ &&&&&&&  \underline{$1$} &  & \\\cline{2-11}
& ${\bf 2}_{-1}$ &&&&&&&   & $1$ & $1$\\\hline
\end{tabular}
\caption{\label{9dTHA}\it Hierarchy of $p$-form fields in $D=9$ as predicted by the Borcherds algebra $\bor_2$. The column $s_p$ lists the $E_2=\mathfrak{sl}(2)\oplus\reals$ representation and the columns with different $q$ represent the various powers that arise in the oxidation process to type IIA supergravity and the entries represent the numbers of such fields. The notation ${\bf n}_h$ for $E_2$ denotes the $n$-dimensional representation of $\mathfrak{sl}(2)$ with eigenvalue $h$ under the $\reals$ summand. The underlined representations are the ones that survive the oxidation process to type IIA (besides the Kaluza--Klein vector). We have truncated the table at $p=6$.}
\end{table}
}
\setlength{\tabcolsep}{6.3pt}
{\renewcommand{\arraystretch}{1.5}
\begin{table}[p]
\begin{tabular}{|c||c||c|c|c|c|c|c|c|c|c|}
\hline
$p$ & $s_p$ & $q=0$ & $q=1$ & $q=2$ & $q=3$ & $q=4$ & $q=5$ & $q=6$ & $q=7$ & $q=8$\\
\hline\hline
\multirow{2}{*}{$1$} & ${\bf 1}_{-4/3}$ &  ${\bf 1}$ &  &  &&&&&&\\\cline{2-11}
& ${\bf 2}_1$ &  &  & ${\bf 2}$ &&&&&&\\\hline
$2$ & ${\bf 2}_{-1/3}$ &&& \underline{${\bf 2}$} &&&&&&\\\hline
$3$ & ${\bf 1}_{2/3}$ &&&& & ${\bf 1}$ &&&&\\\hline
$4$ & ${\bf 1}_{-2/3}$ &&&&& \underline{${\bf 1}$} &&&&\\\hline
$5$ & ${\bf 2}_{1/3}$ &&&&&&   & ${\bf 2}$ &&\\\hline
\multirow{2}{*}{$6$} & ${\bf 2}_{-1}$ &&&&&&&  \underline{${\bf 2}$} && \\\cline{2-11}
& ${\bf 1}_{2/3}$ &&&&&&&  && ${\bf 1}$\\\hline
\end{tabular}
\caption{\label{9dTHB}\it Hierarchy of $p$-form fields in $D=9$ as predicted by the Borcherds algebra $\bor_2$. The column $s_p$ lists the $E_2=\mathfrak{sl}(2)\oplus\reals$ representation and the columns with different $q$ represent the various powers that arise in the oxidation process to type IIB supergravity and the entries represent the representations of the type IIB symmetry algebra $\mathfrak{sl}(2)$.
The underlined representations are the ones that survive the oxidation process to type IIB (besides the Kaluza--Klein vector). We have truncated the table at $p=6$.}
\end{table}
}

\setlength{\arraycolsep}{5.3pt}
{\renewcommand{\arraystretch}{1.5}
\begin{table}[t]
\begin{align*}
\begin{array}{|c||c|c|c|c|c|c|c|}
\hline
\ell_0&\ell_2=0&\ell_2=1&\ell_2=2&\ell_2=3&\ell_2=4&\ell_2=5&\ell_2=6\\
\hline
\hline
\,0\,&{\bf 1}\oplus{\bf 3}\oplus{\bf 1}&\ {\bf 1}\ \quad&&\quad\qquad&&&\\
\hline
1&\dotuline{{\bf 2}}&\dotuline{\underline{{\bf 2}}}&\quad\qquad&&&&\\
\hline
2&&\dotuline{{\bf 1}}&\underline{\bf 1}&\quad\qquad&&&\\
\hline
3&&&\dotuline{{\bf 2}}&\underline{{\bf 2}}&\quad\quad&&\\
\hline
4&&&\dotuline{{\bf 1}}&{\bf 1}\oplus\dotuline{{\bf 3}}&\underline{{\bf 3}}&\quad\qquad&\\
\hline
5&&&&{\bf 2}&2\times{\bf 2}\oplus\dotuline{{\bf 4}}&\underline{{\bf 2}\oplus{\bf 4}}&\\
\hline
6&&&&&{\bf 1}\oplus2\times\dotuline{{\bf 3}}
&2\times{\bf 1}\oplus3\times{\bf 3}\oplus\dotuline{{\bf 5}}&\underline{{\bf 1}\oplus{\bf 3}\oplus{\bf 5}}\\
\hline
\end{array}
\end{align*}
\caption{\label{9dTHAB} \it Level decomposition of $\bor_2$ with respect to the simple roots $\alpha_0$ and $\alpha_2$ (the two black nodes to the right in Figure \ref{B3-dynkin-fig}). The corresponding levels are denoted by $\ell_0$ and $\ell_2$, respectively. For each pair $(\ell_0,\ell_2)$ there
is a representation of the $\mathfrak{sl}(2)$ subalgebra corresponding to the remaining simple root $\alpha_1$ of $\bor_2$.
Whenever $\ell_0=\ell_2$ this representation appears in the $p$-form spectrum of type IIB supergravity, and is therefore underlined in the table. Any positive root of $\bor_2$ is in addition associated to a level $\ell_1$ with respect to the remaining simple root $\alpha_1$. Whenever $\ell_0-\ell_2=\ell_1$ this root is also a root of the type IIA subalgebra and thus contributes to the $p$-form spectrum of type IIA supergravity. Any 
$\mathfrak{sl}(2)$ representation for which such a root occurs (as a weight of the representation) is marked with a dot in the table. The form degree is given by 
$p=\ell_0+\ell_2$ for both type IIA and type IIB. The table is truncated at $\ell_0=6$ and $\ell_2=6$.
}
\end{table}
}

For the decomposition of the $D=9$ representations we have to distinguish two different oxidation processes depending on whether we are aiming for type IIA or type IIB supergravity in $D=10$. We start with the type IIA case, for which Table~\ref{9dTHA} is the relevant one.
We note in particular that there is no four-form that oxidizes from $D=9$ to type IIA supergravity. In the case of type IIB supergravity we are left with an $\mathfrak{sl}(2)$ symmetry algebra in $D=10$. The relevant table for the oxidation of forms is Table~\ref{9dTHB}. 
The $D=9$ representations and their contributions to $p$-form fields in $D=10$ are also summarized in Table~\ref{9dTHAB}, for both type IIA and type IIB.

\section{Further aspects of the Borcherds algebras}
\label{sec:dual}

In this section, we highlight some additional properties of the spectra of the Borcherds algebras $\bor_n$ for $n \geq 3$. 

\subsection{Hodge duality}

In one respect the Borcherds algebra $\bor_n$ is easier to
handle than the Kac--Moody algebra $E_{11}$. Both algebras are infinite-dimensional, and for both algebras the number of irreducible representations at level $p$ increases with $p$. But for $\bor_n$ it grows more slowly, and up to $p=D-3$ it is in fact always equal to one, i.e.~the representations $s_p$ are irreducible. Studying these irreducible representations one finds that $s_p$ is always the conjugate of
$s_{D-2-p}$, which reflects
the Hodge duality between $p$-forms and $(D-2-p)$-forms. This is of course necessary on physical grounds but from an algebraic perspective it is less evident why this has to be true. However, one can show that there is a special structure associated with the analogue of the affine null that also forces this condition in the Borcherds superalgebra. This analogue of the affine null root is the (positive) root associated to the Cartan element that commutes with the $E_{n}$ subalgebra.

The representations $s_p$ of $E_n$ in the level decomposition of $\bor_n$ can be determined up to level $p=D-2$
by studying the corresponding level decomposition of the affine Lie algebra $E_9$. Thus we write
\begin{align}
E_9 &= \bigoplus_{p \in \mathbb{Z}} (E_9)_p, &
(E_9)_0 &= E_n \oplus \mathfrak{sl}(9-n) \oplus \mathbb{R}.
\end{align}
For each $p$, the subspace $(E_9)_p$ is a representation of both $\mathfrak{sl}(9-n)$ and $E_n$. For $p \leq D-3$, the 
representation of $\mathfrak{sl}(9-n)$ is the totally antisymmetric tensor power of $p$ copies of the fundamental representation.
Thus we can use
the result in \cite{Palmkvist:2011vz,Palmkvist:2012nc} which says that $(E_9)_p$ as an $E_n$ representation is the same as 
$s_p$ in the level decomposition of $\bor_n$
for $1 \leq p \leq 9-n$.
Since $E_9$ is the affine extension of $E_8$ we know that its root system consists of all non-zero linear combinations $m\delta + \alpha$ where $\delta$ is the affine null root, $\alpha$ is a root of the $E_8$ subalgebra or zero, and $m$ is an integer. The affine null root corresponds to elements at level $9-n=D-2$ in the level decomposition above. It follows that the representation at level $D-2$ is the adjoint representation of 
$E_n \oplus \mathfrak{sl}(9-n) \oplus \mathbb{R}$, and more generally, that the representation at level $D-2-p$ is the same as at level $-p$, and 
the conjugate of the representation at level $p$, for any $p$. Thus this holds also for the $E_n$ representations $s_p$ in the corresponding
level decomposition of $\bor_n$, for $1 \leq p \leq D-2$.

\subsection{Beyond the space-time limit}
\label{sec:BY}

The Borcherds algebras $\bor_n$ are infinite-dimensional and predict representations of arbitrary `rank' $p$ in the decomposition
(\ref{pdec}), also for $p$ larger than the spacetime dimension $D$. 
Going beyond this spacetime limit, we find that the representations $s_p$ for $2 \leq p \leq D-3$ come back at level $p+(D-2)$, as they do in the $E_9$ (note however that $s_1$ is not included in $s_{D-1}$, and that there is no singlet in $s_{D-2}$). But in addition there are many other representations, and some of them
can be shown to follow a certain pattern up to arbitrary high levels. 
This is done by restricting the root space of $\bor_n$ to that
of $\bor_3$, using the embedding of $\bor_3$ into $\bor_n$ that we have described.
It is straightforward to show that
\begin{align}
\dlb e_0, \dlb e_1,\dlb e_0,\dlb e_3,\dlb e_2,\dlb e_1,  e_0 \drb \drb \drb\drb \drb\drb
\end{align}
is a non-zero element of $\bor_3$. If we keep alternating $(\text{ad }e_0)$ with alternatingly $(\text{ad }e_1)$ and $(\text{ad }e_3)$, we can construct elements at arbitrary high levels with respect to $\alpha_0$.
The $E_3$ elements that these elements belong to can then be `lifted' to $\bor_n$ for $n > 3$ and will have the Dynkin labels
\begin{align} \label{dlodd}
(0, 0, \ldots, 0, 0, \frac{p-3}2, 1)
\end{align}
if $p$ is odd $(p \geq 3)$ and
\begin{align} \label{dleven}
(0, 0, \ldots, 0, 1, \frac{p-4}2, 0)
\end{align}
if $p$ is even ($p \geq 4$).
We recall that the Dynkin labels are the components of the highest weight of the representation in the basis of fundamental weights $\Lambda_1,\,\Lambda_2,\,\ldots,\,\Lambda_n$ (in this order), which is dual to the basis of simple roots, 
$(\Lambda_i,\alpha_j)=\delta_{ij}$. 
The representations given by (\ref{dlodd}) and (\ref{dleven}) will thus `survive' the successive embedding,
from $\bor_8$ down to $\bor_3$. 
This result can also be generalised by involving $(\text{ad }e_2)$ more than once in the construction of the
elements, so that the general Dynkin labels generalising (\ref{dlodd}) and (\ref{dleven}) will involve two parameters instead of only $p$.

We stress that the Dynkin labels (\ref{dlodd}) and (\ref{dleven}) only give some of the irreducible representations
contained in $s_p$ for arbitrary `rank' $p \geq D$ (and also do not count for multiplicities greater than one). 
In order to determine {\it all} the irreducible representations, one can again use the
isomorphism of~\cite{Henneaux:2010ys,Palmkvist:2011vz,Palmkvist:2012nc} that relates the $p$-form representation spaces $s_p$ of $\bor_n$ to representations in a Kac--Moody algebra. Indeed, one can consider $E_{n+p}$ decomposed under $E_n$ to obtain the space $s_p$~\cite{Henneaux:2010ys,Palmkvist:2011vz,Palmkvist:2012nc}. Doing this one finds for example the following four-forms in the $D=3$ hierarchy
\begin{align}
\label{ex3}
p=4:\quad {\bf 6\,696\,000}\oplus {\bf 779\,247} \oplus {\bf 147\,250} \oplus 2\times {\bf 30\,380} \oplus {\bf 3875}\oplus 2\times {\bf 248}
\end{align}
of $E_8$. One can now perform the same analysis as in Table~\ref{3dTH}. When decomposed under $E_7$ the dominant pieces (in terms of the oxidation procedure) are
\begin{align}
\label{ex4}
{\bf 8645}\oplus {\bf 133}
\end{align}
and come from the two largest $E_8$ representations in~(\ref{ex3}). The $E_7$ representations (\ref{ex4}) agree exactly with the four-forms of $D=4$ supergravity~\cite{Riccioni:2007au,Bergshoeff:2007qi,deWit:2008ta,Kleinschmidt:2011vu}. The role of (\ref{ex3}) in $D=3$ is not so clear but some hints might be taken from the point of view of reducibility of constraints in generalised geometry \cite{Berman:2012uy} or from a superspace point of view \cite{Greitz:2011da,Greitz:2012vp}. It has also been observed that the definition of the tensor hierarchy in terms of the embedding tensor predicts an infinite hierarchy~\cite{deWit:2008ta}.

\subsection{Yet more subalgebras}

We finally remark that the Borcherds algebras studied in~\cite{HenryLabordere:2002dk,Henneaux:2010ys} can also be embedded in $\bor_3$, providing a different oxidation scheme than that used in sections~\ref{subsec:8to9} to~\ref{subsec:10to11}. Starting from the diagram of $\bor_3$ 
in Figure~\ref{B3-dynkin-fig} we can consider the following roots
\begin{align}
\label{bor2alt}
\alpha_0,\quad \alpha_0+\alpha_1,\quad \alpha_2.
\end{align}
It is not hard to check that they form a system of simple roots of a subalgebra of $\bor_3$ and that the subalgebra coincides with the one studied in~\cite{HenryLabordere:2002dk}. 
Starting from the alternative Borcherds algebra (\ref{bor2alt}) as the V-duality algebra in $D=9$ one can also recover the other V-duality algebras of~\cite{HenryLabordere:2002dk} that arise in $D>9$ as subalgebras of $\bor_3$. However, the subalgebras correspond to different physical sets of fields that are being kept in the oxidation process and therefore we prefer to study the algebras of sections~\ref{subsec:8to9} to~\ref{subsec:10to11}. 

\subsection*{Acknowledgements}
The authors would like to thank Martin Cederwall, Thibault Damour, Jesper Greitz, Marc Henneaux, Paul Howe, Bernard Julia, Victor Kac and Hermann Nicolai for interesting discussions. JP would like to thank the Albert Einstein Institute for hospitality. Until the end of September 2012, the work by JP was performed at Universit\'e Libre de Bruxelles, and supported by IISN -- Belgium (conventions 4.4511.06 and 4.4514.08) by the Belgian Federal Science Policy Office through the Interuniversity Attraction Pole P6/11.

\bibliographystyle{utphysmod2}


\providecommand{\href}[2]{#2}\begingroup\raggedright\endgroup

\end{document}